\documentclass[letterpaper,aip, amsmath,amssymb,reprint]{revtex4-1}

\usepackage{graphicx}% Include figure files
\usepackage{dcolumn}% Align table columns on decimal point
\usepackage{bm}% bold math
\usepackage[%  
    colorlinks=true,
    linkcolor=blue,
    citecolor=blue,
    urlcolor=blue
]{hyperref}
\usepackage[utf8]{inputenc}
\usepackage[T1]{fontenc}
\usepackage{mathptmx}
\usepackage{etoolbox}
\usepackage{caption}
\usepackage{float}
\usepackage[flushleft]{threeparttable}
\usepackage[title]{appendix}

\graphicspath{ {./Figures/}}

\renewcommand{\arraystretch}{1.25}

\makeatletter
\def\@email#1#2{%
 \endgroup
 \patchcmd{\titleblock@produce}
  {\frontmatter@RRAPformat}
  {\frontmatter@RRAPformat{\produce@RRAP{*#1\href{mailto:#2}{#2}}}\frontmatter@RRAPformat}
  {}{}
}%
\makeatother
\begin{document}

\preprint{AIP/123-QED}

\title{Three dimensional full-field velocity measurements in shock compression experiments using stereo digital image correlation}
% Force line breaks with \\
\author{Suraj Ravindran}
\email[Author to whom correspondence should be addressed: ]{sravi@umn.edu}
    \thanks{These authors contributed equally to this work.}
\affiliation{Aerosapce Engineering and Mechanics, University of Minnesota, Minneapolis, Minnesota 55455, USA}%
\author{Vatsa Gandhi}%
    \thanks{These authors contributed equally to this work.}
\affiliation{Division of Engineering and Applied Science, California Institute of Technology, Pasadena, California 91125, USA}%
\author{Akshay Joshi}
\affiliation{Division of Engineering and Applied Science, California Institute of Technology, Pasadena, California 91125, USA}%
\author{Guruswami Ravichandran}
\affiliation{Division of Engineering and Applied Science, California Institute of Technology, Pasadena, California 91125, USA}%
\affiliation{Jio Institute, Ulwe, Navi Mumbai, Maharashtra 410206, India}

\date{\today}% It is always \today, today,
             %  but any date may be explicitly specified

\begin{abstract}
Shock compression plate impact experiments conventionally rely on point-wise velocimetry measurements based on laser-based interferometric techniques. This study presents an experimental methodology to measure the free surface full-field particle velocity in shock compression experiments using high-speed imaging and three-dimensional (3D) digital image correlation (DIC). The experimental setup has a temporal resolution of $100$ ns with a spatial resolution varying from $90$ to $200$ $\mu$m/pixel. Experiments were conducted under three different plate impact configurations to measure spatially resolved free surface velocity and validate the experimental technique. First, a normal impact experiment was conducted on polycarbonate to measure the macroscopic full-field normal free surface velocity. Second, an isentropic compression experiment on Y-cut quartz-tungsten carbide assembly is performed to measure the particle velocity for experiments involving ramp compression waves. To explore the capability of the technique in multi-axial loading conditions, a pressure shear plate impact experiment was conducted to measure both the normal and transverse free surface velocities under combined normal and shear loading. The velocities measured in the experiments using digital image correlation are validated against previous data obtained from laser interferometry. Numerical simulations were also performed using established material models to compare and validate the experimental velocity profiles for these different impact configurations. The novel ability of the employed experimental setup to measure full-field free surface velocities with high spatial resolutions in shock compression experiments is demonstrated for the first time in this work.\newline\newline
\end{abstract}

\maketitle
\section{Introduction}
Plate impact experiments, generating shock compression, are employed to understand the high pressure and high strain-rate behavior of broad classes of materials, including metals, glasses, ceramics, and composites \cite{meyers_1994}. In these experiments, the free surface velocities are typically measured, which provide insights into various phenomena, such as equation of state\cite{meyers_1994}, inelastic material behavior \cite{abou-sayed_oblique-plate_1976}, polymorphic phase transformations \cite{bancroft_polymorphism_1956}, and spall failure\cite{dandekar_2004}. Laser interferometric techniques are the commonly used method for determining free-surface velocities due to their high sensitivity, accuracy, time resolution, and non-intrusiveness towards propagating shock waves. In these techniques, Doppler-shifted light reflected from the target free surface is superposed with either a delayed version of itself (velocity interferometer) or a reference light beam (displacement interferometer), and the phase difference of the interfered beams is used to obtain the normal and transverse free surface velocities. The most well-known interferometry techniques that are employed for measuring point-wise normal particle velocity are the Velocity Interferometer System for Any Reflector (VISAR) \cite{barker_laser_1972}, and the Photon Doppler Velocimetry (PDV) \cite{strand_compact_2006,mallick2019laser}. The point-wise transverse particle velocities are measured using a transverse displacement interferometer (TDI)\cite{kim_combined_1977} or its corresponding PDV, a heterodyne transverse velocimetry (HTV)\cite{kettenbeil_heterodyne_2018} interferometer. Despite its high temporal resolution ($\sim$ns), common laser interferometry techniques provide only point-wise velocimetry and are unable to capture the spatial heterogeneity in the shock wave. This limits the utility of such techniques in the context of shock compression experiments on heterogeneous materials such as composites, granular, and architected lattice materials, where characterization of the material requires measurement of spatio-temporally resolved free surface velocities. This issue can be partly addressed by employing a variant of VISAR known as Optically Recording Velocity Interferometer System (ORVIS) \cite{bloomquist_optically_1983}, where velocity measurements are conducted along a line to provide improved spatial resolution \cite{vogler_using_2008}. ORVIS requires an elaborate arrangement of optics and high-speed streak cameras in order to provide the desired horizontal scan of the laser needed for the measurement \cite{trott_investigation_2007}. Additionally, this technique only offers insight into the velocity variation along a line on the free surface and not on the entire free surface. Such full-field free surface velocity measurements can be achieved using the Digital Image Correlation (DIC) technique \cite{chu_1985,sutton_image_2009}.

Digital image correlation is well established as a full-field measurement technique for both in-plane and out-of-plane displacement measurements \cite{sutton_image_2009}. This technique involves taking a series of images of a deforming randomly speckled sample and is cross-correlated to extract displacement fields across the entire sample surface \cite{chu_1985,sutton_image_2009}. Therefore, a single experiment yields thousands of displacement measurements over the entire field of view, with sub-pixel accuracy, which can be further used to obtain the velocity, acceleration, and strain fields \cite{pierron_beyond_2014, rubino_full-field_2019}. While DIC is well-suited for quasi-static experiments, advances in high-speed imaging capabilities have enabled DIC in the dynamic regime. For example, high-speed imaging in conjunction with two-dimensional DIC was utilized to study the dynamic micro- to macro-scale deformation behavior of materials \cite{ravindran_local_2016,ravindran_2017, malhotra2022technique, kannan2018mechanics, ravindran_weak-shock_2019}. Recently, Rubino et al. \cite{rubino_full-field_2019} combined high-speed photography, and 2D DIC to characterize deformation fields around dynamically propagating shear ruptures in laboratory earthquake experiments. In addition to in-plane deformation fields, out-of-plane deformation can be resolved using two cameras and high-speed three-dimensional (stereo) digital image correlation, which is used in blast, ballistic and low velocity impact applications. For instance, Tiwari et al. \cite{Tiwari_2009} conducted 3D DIC to measure the deformation of aluminum plates subjected to blast loading, while work by Jannotti et al. \cite{Jannotti_2021} investigated anisotropic ballistic response of rolled AZ31B magnesium plates. Additionally, Gupta et al. \cite{Gupta_2014} conducted 3D DIC to understand the physical processes associated with the implosion of cylindrical tubes submerged underwater. Pankow et al. \cite{Pankow_2010} implemented stereo DIC with a single camera by using a series of mirrors to split the image into two different views and examined the out-of-plane flexural response of a dynamically loaded thin aluminum plate. In these studies, the authors use the extracted full-field displacement and strain field to analyze the deformation mechanisms at time scales ranging from $25-150$ $\mu$s. Overall, previous impact studies using full-field measurements have focused on longer time scale behavior of materials compared to the shock compression behavior where the total event duration is typically smaller than $5$ $\mu$s. 

The present study extends the applications of high speed 3D DIC to plate impact experiments for the first time to understand material behavior in the shock regime (strain rates $>10^6$ s$^{-1}$).  In Section \ref{ExperimentalMethods}, details of the experimental setup, accuracy and resolution of measured data, and the methodology of post-processing the acquired data are discussed. In Section \ref{Results}, the feasibility of the developed experimental setup  for measuring full-field free surface velocity measurements and is first validated against point-wise PDV measurements in normal and pressure-shear plate impact experiments. In Section \ref{Conclusions}, concluding remarks are presented and potential future directions are discussed.

\section{Materials and Methods}\label{ExperimentalMethods}
\subsection{Experimental Setup}
\begin{figure*}[t]
	\centering
	\includegraphics[width=0.9\textwidth]{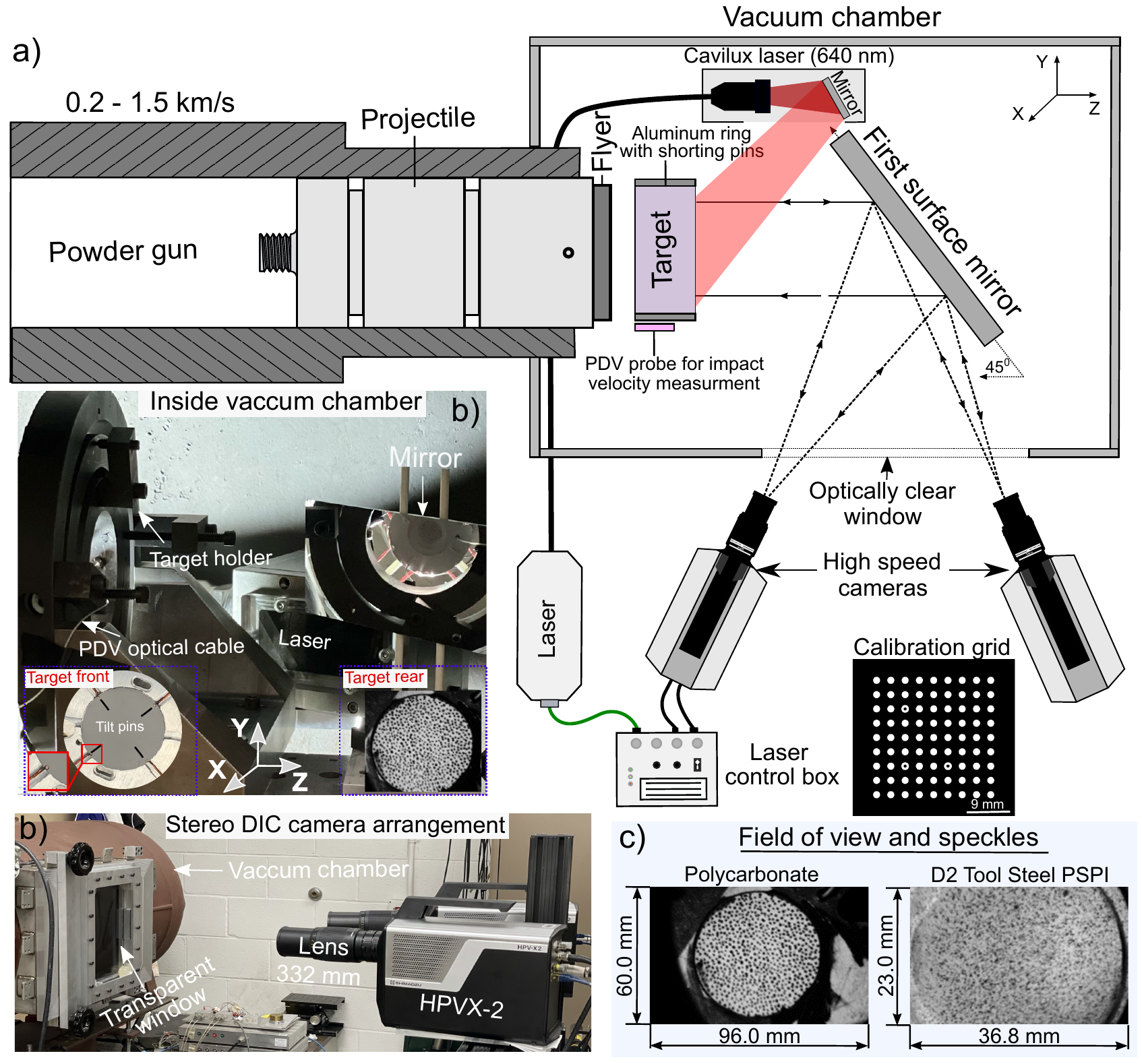}
	\caption{a) Schematic of the experimental setup with powder gun, camera arrangements for 3D digital image correlation, b) image of the components inside the vacuum chamber and the high-speed camera arrangement for stereo DIC set up for imaging through the transparent window, c) representative speckle patterns on the free surface of the sample. }
	\label{fig:ExpSetup}
\end{figure*}

In this study, plate impact experiments were conducted using a powder gun facility at Caltech. Four materials, polycarbonate (PC), tungsten carbide (WC), Y-cut quartz, and D2 tool steel, were investigated in different experimental configurations to demonstrate the reliability, advantage, and robustness of the high-speed stereo (3D) DIC for measuring full-field free surface velocities in plate impact experiments. The schematic of the experimental setup with high-speed imaging-based full-field diagnostics is shown in Fig. \ref{fig:ExpSetup}\hyperref[fig:ExpSetup]{a}. Also, the images of the stereo high speed imaging DIC setup and the components inside the vacuum chamber are shown in Fig. \ref{fig:ExpSetup}\hyperref[fig:ExpSetup]{b}. Samples were shocked using the 38.7 mm slotted powder gun capable of launching projectiles at velocities ranging from 0.2-1.5 km/s. Two Hyper Vision HPV-X2 high-speed cameras (Shimadzu, Kyoto, Japan) capable of capturing 5 million frames/s at full resolution ($400\times250$ pixel$^2$) and 10 million frames/s at half resolution ($400\times250$ pixel$^2$, where missing pixels are interpolated) were used to capture the image of the free surface, see Fig. \ref{fig:ExpSetup}\hyperref[fig:ExpSetup]{a}. These cameras were placed at a stereo angle between $15^{\circ}-16^{\circ}$ and were calibrated by capturing the image of the standard laser engraved calibration grids (Correlated Solutions Inc., Columbia, SC), to facilitate stereo DIC measurement\cite{sutton_image_2009}. A total of 45 image pairs were collected by tilting and rotating the calibration grid within the field of view of both the cameras. It was ensured that the images were within the depth of field of the imaging system to obtain accurate calibration of the stereo camera system.  The image of the camera arrangement and a typical calibration grid used for calibration are shown in Fig. \ref{fig:ExpSetup}\hyperref[fig:ExpSetup]{a}. The images of the rear surface of the sample were observed through a large mirror (50 mm × 100 mm) arranged at $45^{\circ}$ with respect to the loading direction. A CAVILUX Smart laser (Cavitar, Tampere, Finland), which provides high-speed incoherent laser (wavelength, 640 nm) pulses as short as 10 ns, was used to illuminate the sample during loading. The short duration laser pulses avoid blurring due to the transients in the high-speed impact experiments. These laser pulses were synchronized with the exposure of the camera sensor during the image acquisition. An image of the complete target assembly, mirrors, and illumination using laser inside the vacuum chamber is shown in the Fig. \ref{fig:ExpSetup}\hyperref[fig:ExpSetup]{b}. Impact velocities of the experiments were measured using Photonic Doppler velocimetry \cite{strand_compact_2006} (PDV). The PDV probe was attached to the target holder and the optical return was checked during the target alignment.  In the pressure shear plate impact experiment, a heterodyne PDV was used to simultaneously measure normal and transverse velocities\cite{kettenbeil_heterodyne_2018}.

The materials used in this study were obtained from various commercial vendors. Polycarbonate and D2 tool steel materials were procured from McMaster-Carr (Los Angeles, CA) while the tungsten carbide was acquired from Basic Carbide Corporation (Lowber, PA), ARMCO iron from AK Steel International (Breda, The Netherlands), and Y-cut quartz from University Wafer (South Boston, MA). The final dimensions of the flyers and target materials are shown in Table \ref{tab:Overview}.

The flyers and targets were lapped parallel to less than 10 $\mu$m and flat to within 0.5 $\mu$m variation across the surface. Four encapsulated copper shorting pins (tilt pins) were glued to the target and lapped flush to the target surface; see inset in Fig. \ref{fig:ExpSetup}\hyperref[fig:ExpSetup]{b}. These pins were connected to a digital circuit, which was connected to a high-speed digital oscilloscope sampling at 20 GigaSamples/s (MSO9404A, Keysight Technologies, Inc., Santa Rosa, CA) to determine the contact time of the flyer with the target. The contact time of each pin with the flyer was used to calculate the impact tilt in the experiment. The flyer and target assemblies were aligned in the experiment using an auto-collimator and an optically flat mirror assembly prior to the shot\cite{kumar1977optical}. This alignment ensures the impact tilt between the flyer and target below 2 milli-radian in order to generate a plane wave in the sample. A high contrast, random and isotropic speckle pattern was applied on the free surface of the sample to facilitate DIC measurement. The speckles in all the experiments were about $4-6$ pixels in size to avoid anti-aliasing due to low sampling of the pixels \cite{sutton_image_2009}. They were created using a fine tip pen (Pigma Micron, Sakura, Osaka, Japan) for the experiment on polycarbonate. In other experiments, black speckles were applied using an airbrush to obtain higher spatial resolution resulting in smaller speckles (84-92 $\mu$m/pixel). The field of view and the images of the speckle patterns used in the experiments are shown in Fig. \ref{fig:ExpSetup}\hyperref[fig:ExpSetup]{c}. 

\subsection{Postprocessing of images}
 The image sequence acquired using the HPV-X2 high-speed cameras was analyzed to obtain the displacement field using a widely-used commercial software Vic-3D (Correlated Solutions Inc., Columbia, SC). To obtain the stereo camera arrangement calibration parameters, the bundle adjustment algorithm implemented in Vic-3D was used \cite{sutton_image_2009}. This algorithm yields the extrinsic and intrinsic parameters required for calculating all the displacement components from the speckled images acquired during impact. Two important parameters considered during post-processing of the images were the subset size and step size. For a selected subset size, the algorithm calculates all the displacement components at the center of the subset, while the step size determines the distance between the centers of the subsets. For instance, if one chooses step size of 3, the subset calculations will be performed at a distance of 3 pixels from the previous subset and report the data at the center of the subset. Therefore, a step size of 3 will skip three points between the centers of the adjacent subsets. In this study, a significant spatial gradient in displacement was not expected since the goal of this study was to obtain the macroscopic velocity fields and the spatial resolution chosen was much larger than the heterogeneities in the sample. Therefore, subset size of $23\times23$ pixel$^2$ and step size of 7 was used for post-processing. A summary of the experimental resolution and the parameters, such as step and subset size, used for the analysis are shown in Table \ref{tab:DIC}. 
\begin{table}[h]
 	\centering
 	\begin{threeparttable}
 		\setlength{\tabcolsep}{7.5pt}
 		\captionsetup{justification=centering}
 		\caption{Resolution and DIC postprocessing parameters}
 		\centering
 		\footnotesize
 		\begin{tabular}{p{0.1\textwidth}ccc} 
 			\hline \hline
 			\centering\textbf{Material} & \begin{tabular}{@{}c@{}}\textbf{Resolution} \\ \textbf{{[}$\mu$m/pixel{]}}\end{tabular} & \begin{tabular}{@{}c@{}}\textbf{Subset} \\ \textbf{{[}pixel$\times$pixel{]}}\end{tabular} & \begin{tabular}{@{}c@{}}\textbf{Step size} \\ \textbf{{[}pixel{]}}\end{tabular} \\ \hline
 			\centering Polycarbonate           & 200       & $23\times23$            & 7                 \\
 			\centering Quartz-Tungsten Carbide             & 95       & $23\times23$              & 7                 \\
 			\centering D2 Tool Steel          & 93       & $23\times23$             & 7      \\ \hline \hline            
 		\end{tabular}
 		\label{tab:DIC}
 	\end{threeparttable}
 \end{table}

\subsection{Measurement Noise}
\begin{figure*}[t]
	\centering
	\includegraphics[width=1\textwidth]{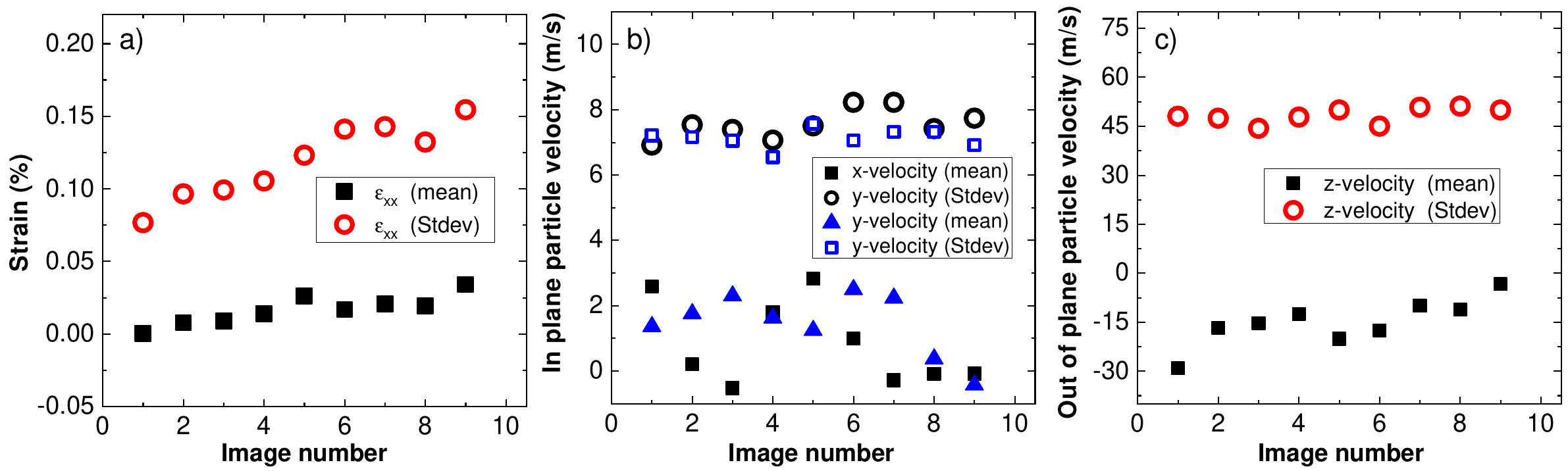}
	\caption{Mean and standard deviation of strain and velocity components from the static images: a) in-plane strain ($\varepsilon_{xx}$), b) in-plane ($x$, $y$) particle velocities, c) free surface (out of plane) $z-$velocity. }
	\label{fig:Noise}
\end{figure*}
All the experiments utilized the half resolution high-speed mode in the camera (Shimadzu HPV-X2) at 10 million frames/second. In the half-resolution mode, every alternate pixel is unused during recording. The grayscale of the unused pixel was interpolated from the neighboring pixels, which produces noise in the measurement. Nine static images were analyzed to quantify the uncertainties associated with the measurement in high-speed mode. The 3D image correlation software, Vic-3D was used to post-process the images to calculate the displacements and strains. Post-processing parameters for each experiment are shown in Table \ref{tab:DIC}. To quantify the velocity uncertainty as a result of random noise in the displacement, the obtained displacement field is numerically differentiated with time increment, $\Delta t = 100$ ns, which is the inter-frame time of the high speed camera at 10 million frames/s.  Figure \ref{fig:Noise} shows the mean and standard deviation of full-field strain and velocity components obtained by post-processing the static images. The highest uncertainty in the strain was around 0.15\% and the uncertainty in the in-plane velocity components was approximately $8$ m/s. At the same time, the out-of-plane velocity uncertainty ($dw/dt=$ $z-$velocity or normal free surface particle velocity) was six times higher than the in-plane free surface particle velocity owing to inherent lower sensitivity in the out-of-plane displacement measurement \cite{sutton_image_2009}.  This higher noise results in 5\% uncertainty when measuring the free surface velocity of 1000 m/s. Note that the uncertainty in velocity was calculated from the raw measurements, which can be significantly reduced by the temporal and spatial filtering of the data\cite{rubino_full-field_2019}. Optimal data filtering is employed to reduce the uncertainties in computing free surface velocities and the procedure is described in Appendix \ref{OptFilt}.

\subsection{Summary of experiments}
The impact configurations, physical dimensions of the flyers, and samples for all the experiments conducted in this study are shown in Table \ref{tab:Overview}. A normal impact experiment on a polycarbonate (PC) sample was performed to show the accuracy and the robustness of the full-field DIC measurement technique in plate impact experiments. The second experiment involved a dual-layer target, tungsten carbide (WC) with a Y-cut Quartz (YQZ) driver, to show the capability of the high-speed DIC measurement to accurately measure the free surface velocity under isentropic compression. Next, a symmetric pressure shear plate impact experiment (PSPI) was conducted on D2 tool steel to evaluate the use of full-field measurements for resolving both normal and transverse full-field velocities from the experiment.
\begin{table*}[t]
 	\centering
 	\begin{threeparttable}
 		\setlength{\tabcolsep}{7.5pt}
 		\captionsetup{justification=centering}
 		\caption{Impact configuration and dimensions of flyer and sample}
 		\centering
 		\footnotesize
 		\begin{tabular}{cccccccc} 
 			\hline \hline
 			\textbf{Experiment}&\begin{tabular}{@{}c@{}}\textbf{Flyer} \\ \textbf{material}\end{tabular}& \begin{tabular}{@{}c@{}}\textbf{Target} \\ \textbf{material}\end{tabular} & \begin{tabular}{@{}c@{}}\textbf{Flyer} \\ \textbf{thickness} \\ \textbf{{[}mm{]}}\end{tabular} & \begin{tabular}{@{}c@{}}\textbf{Target} \\ \textbf{thickness} \\ \textbf{{[}mm{]}}\end{tabular}& \begin{tabular}{@{}c@{}}\textbf{Impact} \\ \textbf{Velocity} \\ \textbf{{[}m/s{]}}\end{tabular} &  \begin{tabular}{@{}c@{}}\textbf{Normal} \\ \textbf{Stress} \\ \textbf{{[}GPa{]}}\end{tabular} & \begin{tabular}{@{}c@{}}\textbf{Tilt} \\ \textbf{{[}mrad{]}}\end{tabular} \\ \hline
 			Normal Impact        & Iron & Polycarbonate (PC) & 4.958$\pm$0.006       & 9.283$\pm$0.001             & 201$\pm$0.5        & 0.55  & 1.8           \\
 			Normal Impact      & WC  & YQZ+WC    & 4.465$\pm$0.002       & \begin{tabular}{@{}c@{}}YQZ --\\ 4.930$\pm$0.005\\ WC -- \\ 1.493$\pm$0.002 \end{tabular}             & 790$\pm$0.1        & 10.4  & 0.9           \\
 			Pressure Shear Plate Impact      &  D2 tool steel & D2 tool steel  & 5.630$\pm$0.004       & 3.983$\pm$0.003             & 530$\pm$0.9        & 10.3   & 1.2         \\ \hline \hline    
        
 		\end{tabular}
 		\label{tab:Overview}
 		\begin{tablenotes}
 			\footnotesize
 			\item **The diameter of the target in normal impact experiment on PC was 28 mm in diameter, diameter of the target in the YQZ+WC was 30 mm, while in pressure shear plate impact experiment, the diameter was 34 mm.
 			\item **The flyer in all experiments was 34 mm in diameter
 		\end{tablenotes}
 	\end{threeparttable}
 \end{table*}
 
 \subsection{Numerical simulation}
 Three-dimensional finite element simulations of the experiments were conducted using the commercial software ABAQUS/Explicit\cite{Abaqus_manual} (Dassault Systemes, Providence, RI) employing previously calibrated material models for PC, Y-cut quartz, WC, and D2 tool steel.  In all the simulations, an element size of $50$ $\mu$m was used based on mesh convergence studies. Further details of the simulations are presented in the respective sections of the experiments. The material models in the simulations include both the equation of state (EOS) and strength models taken from previous studies. The volumetric part of the deformation (Pressure-Volume Hugoniot relation) was modeled using linear Gr\"{u}niesen equation of state\cite{meyers_1994}:
 
\begin{equation}
P_{H} = \frac{\rho_0C_0 \upsilon}{(1-S\upsilon)^2}
\label{eq:Hugoniot}
\end{equation}

\noindent where, $C_0$ is the bulk sound speed, $S$ is the slope of linear shock velocity ($U_s$) - particle velocity ($u_p$) relation, $U_s = C_0 + Su_p$, $u_p$ is the normal particle velocity, $\upsilon$ is the volumetric strain, $\upsilon = 1- \frac{\rho_0}{\rho} = \frac{u_p}{U_s}$, where $\rho_0$ and $\rho$ are the initial and final densities, respectively. In the simulations, parameters ($C_0$, $S$)  of the linear equation of state were taken from past studies and are shown in Table \ref{tab:EOS}.
 
\begin{table}[h]
 	\centering
 		\setlength{\tabcolsep}{7pt}
 		\captionsetup{justification=centering}
 		\caption{Equation of state parameters for the materials in the experiments.}
 		\centering
 		\footnotesize
 		\begin{tabular}{cccc} 
 			\hline \hline
 			\centering\textbf{Material} & \begin{tabular}{@{}c@{}}\textbf{Density, $\rho_0$} \\ \textbf{{[}kg/m$^3${]}}\end{tabular} & \begin{tabular}{@{}c@{}}\textbf{Bulk Sound} \\ \textbf{Speed, $C_0$}\\ \textbf{{[}m/s{]}}\end{tabular} & $S$ \\ \hline
 			Iron\cite{barker_shock_1974} & 7800 & 4460 & 1.72 \\
 			Polycarbonate \cite{dwivedi_mechanical_2012,carter_hugoniot_1995}           & 1197       & 2330            & 1.57                 \\
 			Tungsten Carbide\cite{Dandekar_2002}             & 15600      & 4930              & 1.30                 \\
 			D2 Tool Steel \cite{Ravindran_2021}          & 7900       & 4590             & 1.42      \\ \hline \hline            
 		\end{tabular}
 		\label{tab:EOS}
 \end{table}

Two different models were used to capture the strength behavior of the materials used in the experiments. The flow strength $(Y)$ of polycarbonate (PC) was modeled using the Johnson-Cook material model with the calibration parameters obtained from the Dwivedi et al. \cite{dwivedi_mechanical_2012} and Rai et al.\cite{rai_mechanics_2020}, respectively.

\begin{equation}
Y=\left(A+B \varepsilon_{p}^{n}\right)\left(1+C \ln \frac{\dot{\varepsilon}}{\dot{\varepsilon}_{r e f}}\right)\left(1-\left(\frac{T-T_{R}}{T_{m}-T_{R}}\right)^{m}\right)
\label{eq:JC}
\end{equation}

\noindent where $A, B$, and $n$ are the parameters for the power-law strain hardening, $\varepsilon_p$ is the plastic strain, and $C$ is the constant for the strain rate hardening, $\dot{\varepsilon}_{ref}$ is the reference strain rate, and $m$ is the exponent for thermal softening. Tungsten carbide and D2 tool steel were modeled using a modified Johnson-Cook model, in which the logarithmic strain rate hardening term is replaced with the Cowper-Symmonds power-law strain rate hardening:

\begin{equation}
Y=\left(A+B \varepsilon_{p}^{n}\right)\left(1+\left(\frac{\dot{\varepsilon}_{p}}{D \dot{\varepsilon}_{r e f}}\right)^{p}\right)\left(1-\left(\frac{T-T_{R}}{T_{m}-T_{R}}\right)^{m}\right)
\label{eq:CowperSym}
\end{equation}

\noindent where $p$ and $D$ are parameters associated with the strain-rate hardening term. Parameters for these models were obtained from the past studies and listed in Table \ref{tab:MatParam}. Additionally, thermal softening was not considered in the simulations as the estimated temperatures under the loading conditions considered here is only a small fraction of the melting temperature.
 
\begin{table*}[t]
    \setlength{\tabcolsep}{7.5pt}
    \captionsetup{justification=centering}
 	\caption{Parameters of the strength model used in simulations for Polycarbonate (PC), Tungsten Carbide (WC), Iron, and D2 tool Steel. }
 	\centering
 	\footnotesize
    \begin{tabular}{cccccccccc}
    \hline\hline
    \multicolumn{1}{l}{} & \multicolumn{1}{l}{}                  & \multicolumn{8}{c}{Model parameters (Eqs. (\ref{eq:JC}) and (\ref{eq:CowperSym}))}                                                                                                                         \\ \cline{3-10} 
    \textbf{Material}    & \begin{tabular}{@{}c@{}} \textbf{Shear Modulus} \\ $G$ \textbf{{[}GPa{]}}\end{tabular} & \begin{tabular}{@{}c@{}} $A$ \\ \textbf{{[}MPa{]}}\end{tabular} & \begin{tabular}{@{}c@{}} $B$ \\ \textbf{{[}MPa{]}}\end{tabular} & \textbf{$n$} & \textbf{$C$} & \textbf{$m$} & \textbf{$p$} & \begin{tabular}{@{}c@{}} $D$ \\ ($\times10^3$)\end{tabular} & \textbf{$\dot{\varepsilon}_{ref}$} \\ \hline
    PC\cite{dwivedi_mechanical_2012}                   & 0.93                                  & 80                     & 70                     & 2.00         & 0.052        & 0.548        & -            & -                           & 1                                  \\
    WC\cite{Ravindran_2021}                   & 278                                   & 4300                   & 9200                   & 0.38         & -            & -            & 2            & 1500                        & 1                                  \\
    D2 tool steel \cite{Ravindran_2021}               & 78                                  & 850                    & 150                    & 0.24         & -            & -            & 2            & 500                         & 1                                  \\
    Iron\cite{sadjadpour_2015}                 & 78                                   & 32.6                    & 430                     & 0.1         & 0.28       & 0.55         & -            & -                           & 355        \\ \hline\hline                         
    \end{tabular}
    \label{tab:MatParam}
\end{table*}

Unlike other materials, Y-cut quartz (YQZ) was modeled using the anisotropic elastic model since the pressures that we are interested in the study do not create appreciable plastic strain in the samples. Anisotropic elastic constants and the density of the YQZ were obtained from Johnson et al.\cite{johnson_shock_1971} and  Heyliger et al. \cite{Heyliger_2003} and listed in Table \ref{tab:quartz}.

\begin{table}[h]
 	\centering
 		\setlength{\tabcolsep}{7pt}
 		\captionsetup{justification=centering}
 		\caption{Density and elastic constants of Y-cut quartz \cite{johnson_shock_1971,Heyliger_2003}}
 		\centering
 		\footnotesize
 		\begin{tabular}{ccccccc} 
 			\hline \hline
 			\begin{tabular}{@{}c@{}} Density \\ {[}kg/m$^3${]}\end{tabular} & \begin{tabular}{@{}c@{}} $C_{11}$ \\ {[}GPa{]}\end{tabular} & \begin{tabular}{@{}c@{}} $C_{12}$ \\ {[}GPa{]}\end{tabular}& \begin{tabular}{@{}c@{}} $C_{13}$ \\ {[}GPa{]}\end{tabular} & \begin{tabular}{@{}c@{}} $C_{14}$ \\ {[}GPa{]}\end{tabular}& \begin{tabular}{@{}c@{}} $C_{33}$ \\ {[}GPa{]}\end{tabular} & \begin{tabular}{@{}c@{}} $C_{44}$ \\ {[}GPa{]}\end{tabular} \\ \hline
 			2650         & 86.8      & 7.0             & 11.9 & -18.0 & 105.8 & 58.2      \\ \hline \hline            
 		\end{tabular}
 		\label{tab:quartz}
 \end{table}

\section{RESULTS AND DISCUSSION} \label{Results}

\subsection{Normal impact experiment on polycarbonate}\label{Polycarbonate}
\begin{figure*}[t]
	\centering
	\includegraphics[width=0.9\textwidth]{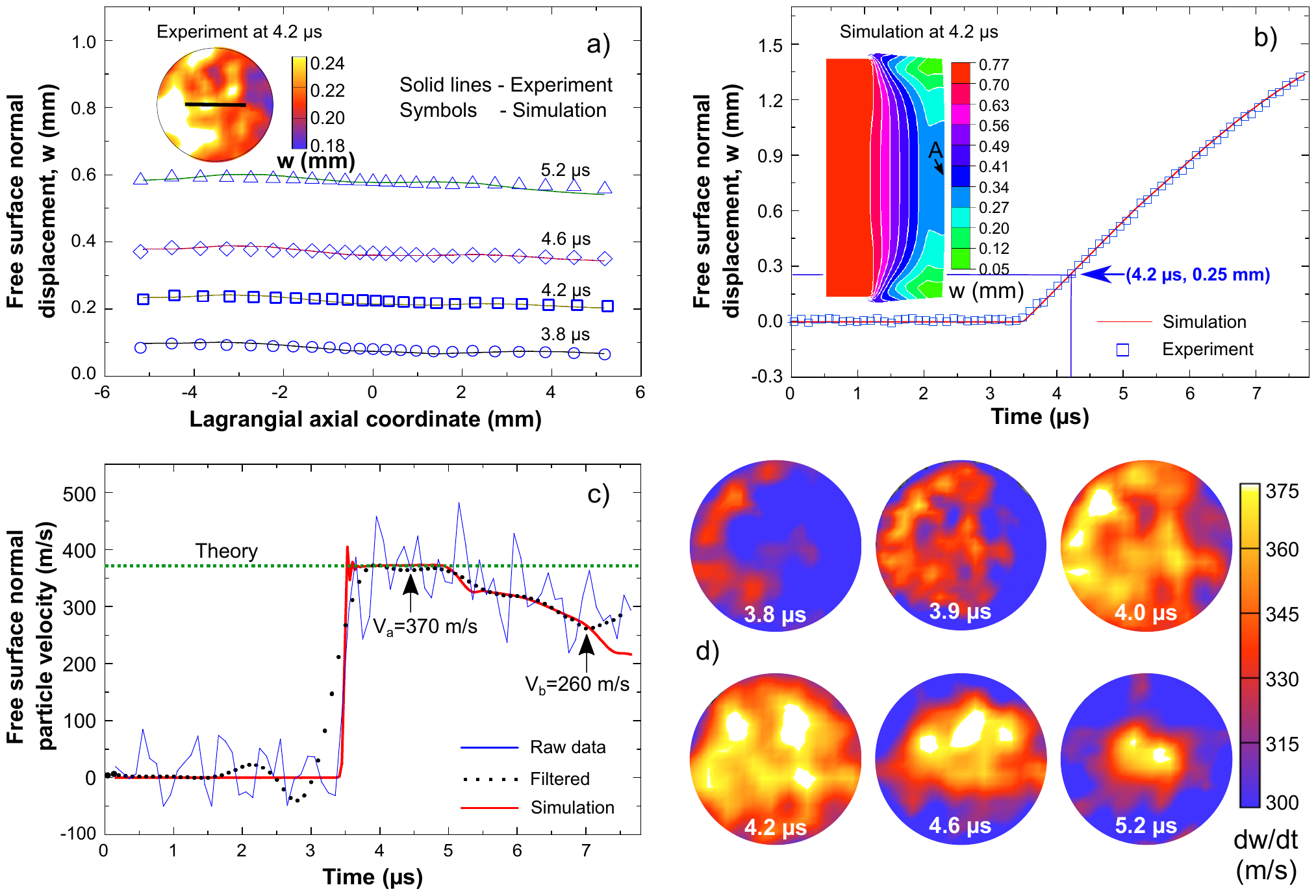}
	\caption{a) Variation of the free surface displacement $(w)$ from experiment and simulation in a normal plate impact on polycarbonate at different Lagrangian axial coordinates along the line shown in the inset. The inset figure shows the full-field particle displacement at $t=4.2$  $\mu$m, b) evolution of free surface displacement from simulation and experiment at the center of the sample, where the inset shows the displacement from the simulation, c) normal free surface particle velocity $(d w / d t)$ from simulation and experiment at the center of the target, d) full-field free surface particle velocity at different times.}
	\label{fig:PC}
\end{figure*}

A pure iron flyer impacted a polycarbonate target at a velocity of $201 \pm 0.5 $ m/s in a normal plate impact experiment. The stress associated with the impact calculated using impedance matching \cite{meyers_1994} was $0.58$ GPa. For these calculations, linear equation of state (EOS) of iron and polycarbonate was used, see Table \ref{tab:EOS}. Images of the speckled free surface taken using a high-speed camera during the experiment were used to calculate the full-field free surface displacement of the sample. Figure \ref{fig:PC}\hyperref[fig:PC]{a} (inset) shows the full field out-of-plane (free surface) displacement ($w$) at $t = 4.2$ $\mu$s, with $t=0$ being the time of impact. It is noted that the displacement at the left edge is about 60 $\mu$m higher than at the right edge of the sample. To visualize the variation of the displacement at a given time, out of plane displacement at four different times, $t=3.8$, 4.2, 4.6, and 5.2 $\mu$s, along a 9 mm line that passes through the center of the sample is plotted in Fig. \ref{fig:PC}\hyperref[fig:PC]{a}. The observed spatial displacement variation along the line is associated with the impact tilt in the experiment. The average tilt measured using the shorting pins was $1.8$ mrad. Numerical simulations were performed to evaluate the accuracy of the free surface displacement measurement and other derived quantities. In the 3D simulation, the flyer was tilted in the same orientation as in the experiment, which was measured using tilt pins, to simulate the exact impact conditions of the experiment. Material strength and the pressure-volume behavior were modeled using the Johnson-cook material model and the linear equation of state $\left(U_{s}-u_{p}\right)$, respectively. Parameters for the linear equation state and the Johnson-cook model are shown in Tables \ref{tab:EOS} and \ref{tab:MatParam} respectively. It is noted that the spatial variation of the displacement in the experiment (solid lines) and simulation (symbols) closely match with each other, including the higher displacement at the left edge, as shown in Fig. \ref{fig:PC}\hyperref[fig:PC]{a}. Also, the temporal evolution of the free surface displacement at the center of the sample matches closely with the simulation, see Fig. \ref{fig:PC}\hyperref[fig:PC]{b}. The inset in Fig. \ref{fig:PC}\hyperref[fig:PC]{b} shows the section at the center of the sample from the simulation at $t = 4.2$ $\mu$s, which indicates that the displacement at the center is in close agreement to the displacement measured ($0.25$ mm) in the experiment.

The particle velocity was calculated by numerically differentiating the displacement data using a central difference formula. Figure \ref{fig:PC}\hyperref[fig:PC]{c} shows the particle velocity calculated from the raw displacement data, filtered displacement data, and simulation. Calculation of the velocity from noisy displacement data shows unrealistically large oscillations with magnitude close to $60$ $m/s$. Hence, the random noise in the displacement data was filtered using the method described in Appendix \ref{OptFilt}. Consequently, the velocity calculated from the filtered data shows nearly constant steady-state free surface velocity, which is close to the simulated and theoretically calculated particle velocity. Filtering reduces the uncertainty in the out-of-plane velocity down to $20$ m/s. However, the rise time of the experiment is difficult to capture due to the reduced temporal resolution from filtering. The evolution of full-field free surface particle velocity calculated from the full-field displacement is shown in Fig. \ref{fig:PC}\hyperref[fig:PC]{d}. These particle velocity contours also show the effect of tilt on the spatial distribution; at $t=3.8$ $\mu$s, the velocity at the left edge is higher than the sample's right edge. However, after $t=4.0$ $\mu$s, the free surface shows a relatively homogeneous velocity field. It indicates that the flyer takes about 200 ns to close the gap with the target, which corroborates with the tilt measured in the experiment.

Shock wave velocity was also calculated using the known thickness of the sample and shock wave transit time in the sample. Shock transit time was calculated by finding the time difference between the contact time of the flyer with the sample and the wave arrival time at the free surface. The contact time was obtained from the copper shorting pins, while the arrival time of the shock at the free surface was determined by inspecting the temporal evolution of the out-of-plane displacement or particle velocity. Shock arrival time at the free surface of the sample was $3.5 \pm 0.1$ $\mu$s. Thus, the shock velocity in the sample was calculated (sample thickness/time of shock arrival) to be $2620 \pm 75$ m/s, which is in accord with the previously reported shock velocity of 2553 m/s at similar pressures\cite{ye_spallation_2019}. 

The normal stress ($\sigma$) from the shock wave velocity and particle velocity can be calculated using\cite{meyers_1994},

\begin{align}
\sigma=\rho_{0} U_{s} u_{p}
\label{eq:stress}
\end{align}

\noindent where, $\rho_{0}$ is the initial density, $U_{s}$ is the shock velocity, which is calculated from the linear equation of state, $U_{s}=C_{0}+S u_{p}$ of the sample, $S$ is the slope of the linear $U_s-u_p$ relation, $C_{0}$ is the bulk sound velocity of the sample, and $u_{p}$ is the particle velocity, which is half the measured normal free surface particle velocity ($dw/dt$). The shock stress calculated from the unfiltered data using Eq. (\ref{eq:stress}) was $0.58 \pm 0.12$ GPa, where the uncertainty is associated with both the particle velocity and shock wave velocity and is around $20 \%$. The uncertainty in the particle velocity can be reduced by removing the random noise in the measured displacement, as discussed in the Appendix. After filtering the displacement, the stress in the sample was calculated to be $0.58 \pm 0.06$ GPa, with the uncertainty reduced to $10 \%$.

The experiment was also designed to measure the spall strength \cite{meyers_1994}. Free surface velocity histories obtained at the center of the sample is used to deduce the spall strength of the material,

\begin{align}
\sigma_{\text {spall }}=\rho_{0} C_{L} \Delta u \frac{1}{1+\frac{C_{L}}{C_{B}}}  
\label{eq:Spall}
\end{align}

\noindent where $C_{L}$ and $C_{B}$ are the longitudinal wave speed and bulk sound speed, respectively, and $\Delta u$ is the pullback velocity determined from the free surface velocity measurement. The values of the various parameters used in the calculation of spall strength using Eq. (\ref{eq:Spall}) are, $C_{L}=2190$ m/s, $C_{B}=1930$ m/s, $\Delta u=V_{a}-V_{b}=110$ m/s, see Fig. \ref{fig:PC}\hyperref[fig:PC]{c}. The spall strength of polycarbonate estimated from this experiment at a normal stress of $0.58$ GPa and a strain rate of $0.12 \times 10^{5}$ s$^{-1}$ is $0.14$ GPa. The spall strength reported for polycarbonate at a similar shock stress (at a higher strain rate) is $0.16$ GPa \cite{majewski1986tension}. This is $12.5$ \% higher than the measured value in this study and the discrepancy could be attributed to strain rate effects and measurement uncertainty. 
\begin{figure*}[t]
	\centering
	\includegraphics[width=0.8\textwidth]{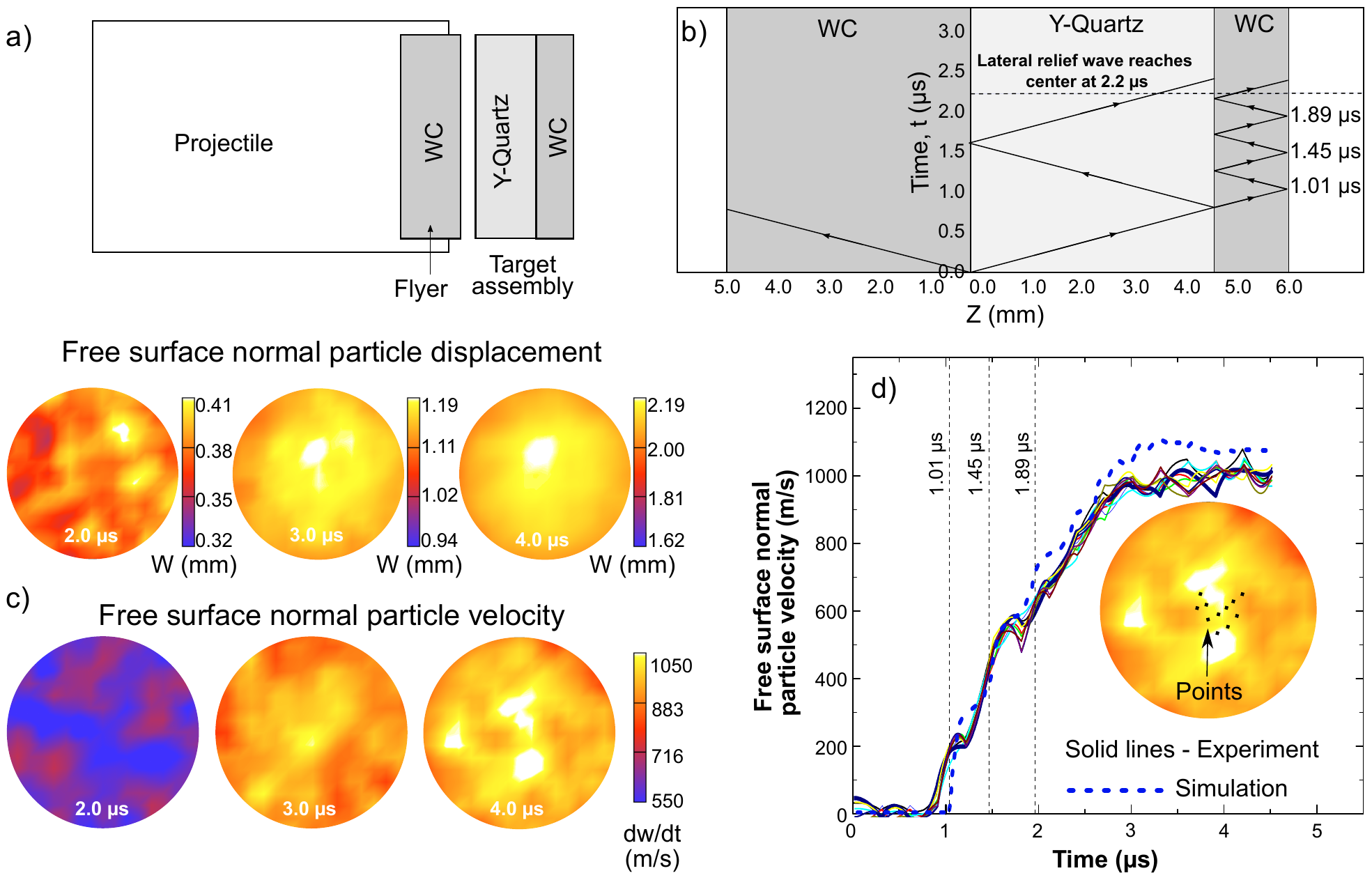}
	\caption{ a) Experimental configuration with $\mathrm{Y}$-cut quartz-tungsten carbide $(\mathrm{YQZ}+\mathrm{WC})$ target assembly, b) distance-time $(t-Z)$ diagram of the experimental configuration, c) free surface particle displacement $(w)$ and free surface particle velocity $(d w / d t)$ at $t=2.0$ $\mu$s, $3.0$ $\mu$s, and $4.0$ $\mu$s, d) free surface particle velocity evolution at different points marked on the free surface of the WC. }
	\label{fig:YQZ}
\end{figure*}

\subsection{Normal impact experiment on Quartz-WC assembly}

The high-pressure properties of condensed matter under isentropic compression is important in many dynamic applications. In such loading, the rise time is relatively large compared to loading conditions that generate strong shocks. A model experiment was conducted to demonstrate the capability of high-speed DIC experiments for measuring the free surface velocities with long rise times. To resolve the longer rise times ($>300$ ns), measurement diagnostics with a time resolution of 100 ns is deemed sufficient. Y-cut quartz (YQZ)-tungsten carbide (WC) dual-layer assembly was used as a target in this experiment, see Fig. \ref{fig:YQZ}\hyperref[fig:YQZ]{a}. The driver was the YQZ crystal of approximately 5 mm in thickness and 30 mm in diameter, and the measurement was performed on the rear side of the WC of the target, see Fig. \ref{fig:YQZ}\hyperref[fig:YQZ]{a}. The flyer material was also made of tungsten carbide. Figure \ref{fig:YQZ}\hyperref[fig:YQZ]{b} plots the distance-time $(t-Z)$ diagram of the experiment, which shows the wave reverberation within the thin WC attached to the YQZ. Three recompression signals can be recorded for a total duration of $1.89$ $\mu$s before the lateral relief wave arrives at the center of the sample. This long rise time of $1.89$ $\mu$s allows to capture 18 images for the displacement measurement; therefore, it was relatively easy to resolve the temporal features of the free surface velocity history.

Figure \ref{fig:YQZ}\hyperref[fig:YQZ]{c} shows the full-field free surface displacement and the free surface particle velocity at different times $t=$\( 2.0\ \mu \)s, \( 3.0\ \mu \)s, and \( 4.0\ \mu \)s. The displacement at various times is relatively homogeneous, with the peak displacements of \( 410\ \mu \)m, \( 1190\ \mu \)m, and \( 2190\ \mu \)m at $t=$\( 2.0\ \mu \)s, \( 3.0\ \mu \)s, and \( 4.0\ \mu \)s, respectively. The peak out-of-plane displacement (or normal displacement, $w$) was observed to be near the center of the WC as a result of the uniaxial strain condition in shock compression. The normal free surface velocity $(d w / d t)$ contours determined from the full-field displacement is also plotted in Fig. \ref{fig:YQZ}\hyperref[fig:YQZ]{c}, showing a homogeneous field. The small heterogeneity in the velocity is mainly associated with the noise in the measurement, similar to the observation in the normal impact experiments on polycarbonate described earlier in Section \ref{Polycarbonate}. To visualize the free surface velocity variation as a function of time, the data was plotted at randomly selected points shown in the inset of Fig. \ref{fig:YQZ}\hyperref[fig:YQZ]{d}. The velocity evolution at different points on the free surface indicates that the magnitude of free surface velocities is very close to each other since the microscale heterogeneity in WC is at much smaller length scales. The temporal evolution of the particle velocity shows step-like features at $t=1.01$ $\mu$s, 1.45 $\mu$s, and 1.89 $\mu$s, see Fig. \ref{fig:YQZ}\hyperref[fig:YQZ]{d}. This timing corresponds to the arrival of the reflected waves from the YQZ-WC interface, depicted in Fig. \ref{fig:YQZ}\hyperref[fig:YQZ]{b}, indicating that these step-like features are due to wave reverberations with the WC target.

Numerical simulations were conducted using the material models discussed in Section \ref{Polycarbonate}, and the resulting free surface velocity is plotted in Fig. \ref{fig:YQZ}\hyperref[fig:YQZ]{d}. The Y-cut quartz was modeled using an anisotropic elastic constitutive description, and the material parameters are shown in Table \ref{tab:quartz}. WC was modeled using the linear equation of state and the strength model with constitutive description in Eq. (\ref{eq:CowperSym}). The strength model was calibrated in our previous study using the data from symmetric pressure shear plate impact (PSPI) experiments\cite{Ravindran_2021}, and the parameters are shown in Table \ref{tab:MatParam}. Figure \ref{fig:YQZ}\hyperref[fig:YQZ]{d} shows the normal particle velocity calculated from the numerical simulations and the experiments. Notably, the step-like features resolved in the experiments were also reproduced by the simulations, indicating the confidence of the measurement technique developed in this study. Hence, the temporal resolution used in the study is sufficient to capture the longer rise time measurements with different temporal features seen in Fig. \ref{fig:YQZ}.

\subsection{Pressure shear plate impact experiment on D2 tool steel}
\begin{figure*}[t]
	\centering
	\includegraphics[width=0.9\textwidth]{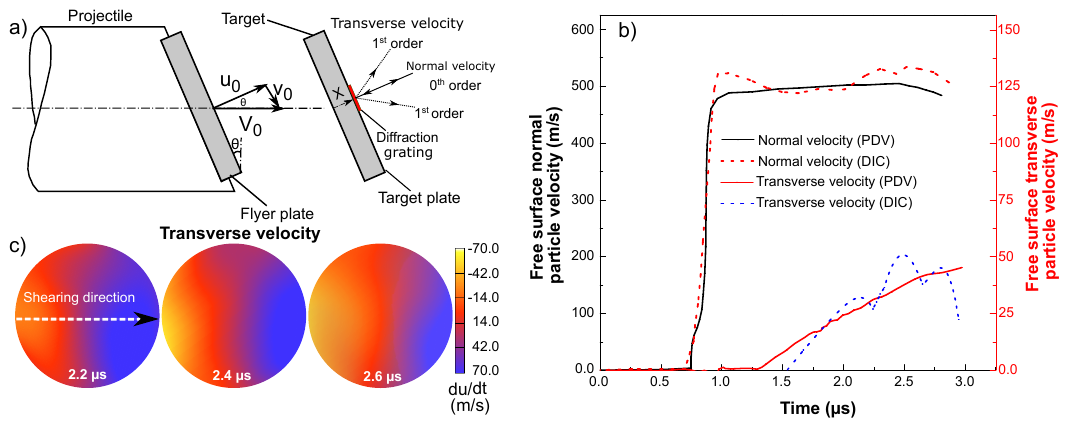}
	\caption{a) Pressure shear plate impact (PSPI) configuration, b) free surface normal and transverse particle velocity profiles for D2 tool steel from PDV/HTV and high-speed DIC measurements, c) contour plots of transverse velocity $(d v / d t)$ at $t=2.2$ $\mu$s, $2.4$ $\mu$s, and $2.6$ $\mu$s. }
	\label{fig:PSPI}
\end{figure*}
Pressure shear plate impact (PSPI) experiments are advantageous in measuring the shear strength of materials at high pressures and strain rates\cite{abou-sayed_oblique-plate_1976,Ravindran_2021,kettenbeil_pressure-shear_2020,Klopp_1985}. The difference between PSPI and normal plate impact experiments is the skewness $(\theta)$ of the flyer plate and target plate configuration, as shown in Fig. \ref{fig:PSPI}\hyperref[fig:PSPI]{a}. The inclined impact generates both longitudinal and shear waves within the flyer and the target assembly. In these experiments, normal velocities are typically measured using photonic Doppler velocimetry (PDV) and transverse velocity using heterodyne transverse velocimeter\cite{kettenbeil_heterodyne_2018} (HTV), transverse displacement interferometer\cite{kim_combined_1977} (TDI), or modified PDV \cite{mallick2019simple}. This is done by depositing a diffraction grating on the free surface of the target where the diffracted $0^{\textnormal{th}}$ order beam is interfered with a reference signal to extract the normal free surface velocity and the diffracted $\pm1^{\textnormal{st}}$ order beams are interfered with each other to determine the transverse velocity. Detailed descriptions of PSPI experiments and the velocity measurement techniques can be found elsewhere \cite{kettenbeil_heterodyne_2018,kettenbeil_pressure-shear_2020,Klopp_1985,kim_combined_1977, zuanetti2017compact}. For PSPI experiments, in addition to the fabrication of the diffraction grating, the main difficulty arises from the precise alignments required to ensure that (1) the diffraction plane is perfectly aligned with the shearing direction, and (2) the fiber optic probes are precisely aligned to maximize the light input in the transverse probes. Also, the transverse signals are highly sensitive to impact tilt which could result in light loss during an experiment. On the other hand, full-field DIC does not require precise alignment and it is insensitive to impact tilt, therefore, using DIC relaxes the tight constraints required in conventional (PDV, HTV) transverse velocity measurements. Thus, digital image correlation can be a powerful free surface velocity measurement technique for PSPI experiments. To explore its feasibility, a PSPI experiment was conducted on D2 tool steel with an impact skew angle of $\theta=18^{\circ}$ and the full-field normal and transverse velocities were measured using high-speed digital image correlation. This measurement was then compared with our previous PSPI experiment\cite{Ravindran_2020} with similar loading conditions that utilized laser interferometry (PDV/HTV) to extract the normal and transverse free surface velocities at a single point, i.e., the center of the rear free surface. 

The impact velocity of the D2 tool steel flyer plate in the experiment was measured to be $530 \pm 0.9$  m/s , which was close to a previous PSPI experiment ($539 \pm 0.8$ m/s) with PDV and HTV based single point velocity measurements\cite{Ravindran_2020}. Figure \ref{fig:PSPI}\hyperref[fig:PSPI]{b} shows the normal and transverse free surface velocity at the center of the sample from high-speed DIC measurement along with the PDV-HTV measurements from an earlier study\cite{Ravindran_2021}. As expected, the normal and transverse components of the impact velocity match the measured free surface velocities of both the DIC and the laser interferometry results. Comparing the DIC data to the PDV measurements, the normal velocity arrival time and the final Hugoniot state of the material were close, considering the uncertainty associated with the out-of-plane velocity measurement (Fig. \ref{fig:Noise}) and the temporal resolution of the experiment. Also, the high-speed DIC measurement captures the shock wave rise time relatively well, however, resolving the Hugoniot elastic limit was difficult due to the low temporal resolution in the DIC measurement. The peak transverse velocity that was measured in the previous experiment, approximately 45 m/s, closely matched the transverse velocity profile measured in high-speed DIC measurements. However, the DIC transverse velocity profile shows oscillatory behavior, which is associated with uncertainty due to numerical differentiation of the noisy transverse displacement. The shear wave dissipates when traveling through the plastically deformed medium due to the forerunning normal shock which reduces the transverse displacement magnitude on the backside of the target. The displacement resolution required is high to resolve such small displacements, which can be increased by using higher magnification optics. 

Figure \ref{fig:PSPI}\hyperref[fig:PSPI]{c} shows the full-field transverse particle velocity from the experiment. Note that the transverse velocity at the edge of the sample is higher than $70$ m/s due to the lateral deformation at the circumference of the target sample. At the center of the sample, the velocity is very close to the previously measured transverse velocity in this material in the PSPI configuration using single point laser interferometry.
\section{Summary and conclusion}\label{Conclusions}

An experimental methodology has been developed for the first time to measure the full-field particle velocity in shock compression experiments using high-speed imaging and stereo 3D digital image correlation (DIC). Since DIC cross-correlates speckle images to determine displacement fields, computing the derivative from the data with noise to obtain velocities is challenging. To overcome this limitation, an optimal smoothing method (See Appendix \ref{OptFilt}) was adopted to filter the data such that the roughness of the data is minimized without oversmoothing the key features in particle velocity. Three different types of experiments were conducted to validate the measurement technique to resolve free-surface velocities in experimental configurations involving shock compression, which are summarized below.
\begin{enumerate}
    \item A normal impact experiment on polycarbonate at low spatial resolution demonstrated that the full-field free surface velocity under shock compression can be measured using high-speed digital image correlation. The results from numerical simulation closely matched the filtered free surface velocity using the optimal spline smoothing of the data. The temporal resolution of the experiment was critical to resolve the shock rise time, which can be challenging in strong shock experiments, where the shock thicknesses are relatively small.
    \item In the normal impact experiment on a Y-cut quartz-tungsten carbide (YQZ-WC) target assembly, the long rise time experimental profile associated with isentropic compression was captured remarkably well and matched the numerical simulation. In this experiment, the spatial resolution used was around 94 $\mu$m, a factor of two smaller than in the polycarbonate experiment. Hence, the velocity uncertainty was reduced to 30 m/s as compared to 45 m/s uncertainty for the polycarbonate experiment.
    \item A symmetric pressure shear plate impact (PSPI) experiment was conducted on D2 tool steel to show the capability of the technique in measuring both normal and transverse velocities under combined normal and shear loading arising from an oblique impact. The high-speed DIC measurement closely matched the previously measured normal and transverse velocity using point-wise measurement using PDV and HTV interferometric techniques.
\end{enumerate}

The newly developed technique offers many advantages for both normal and oblique plate impact experiments. DIC requires cameras and a speckled sample, which circumvents traditional interferometer setups involving complicated optical arrangements, laser alignments, and for oblique impacts, diffraction gratings \cite{kettenbeil_heterodyne_2018}. The high-speed DIC technique is convenient for measuring the spatially resolved data. With the current state-of-the-art cameras, the best time resolution that can be used for accurate 3D digital image correlation measurement is $100$ ns, which is sufficient to resolve rise time as short as 300 ns. The spatial resolution can be achieved as small as $0.2$ $\mu$m/pixel, enabling the measurement of the local particle velocities in heterogeneous materials with grain sizes as small as $5$ $\mu$m. This capability allows one to probe material behavior at mesoscale under shock compression, which remains a great challenge with conventional optics based techniques. It is expected that the framing rate of cameras will increase in the future, enabling one to measure the full field particle velocity in strong shock regimes where the rise time are below 10 ns.

\begin{acknowledgements}
The research reported here was supported by the DOE/NNSA (Award No. DE-NA0003957), which is gratefully acknowledged. The authors acknowledge the support from the Office of Naval Research (Award No. N00014-16-1-2839) for developing the high-pressure PSPI capability, and the Army Research Laboratory (Cooperative Agreement Number W911NF-12-2-0022) for the acquisition of the high-speed cameras.
\end{acknowledgements}

\appendix

\section{Optimal data filtering}\label{OptFilt}

\begin{figure*}[ht]
	\centering
	\includegraphics[width=0.8\textwidth]{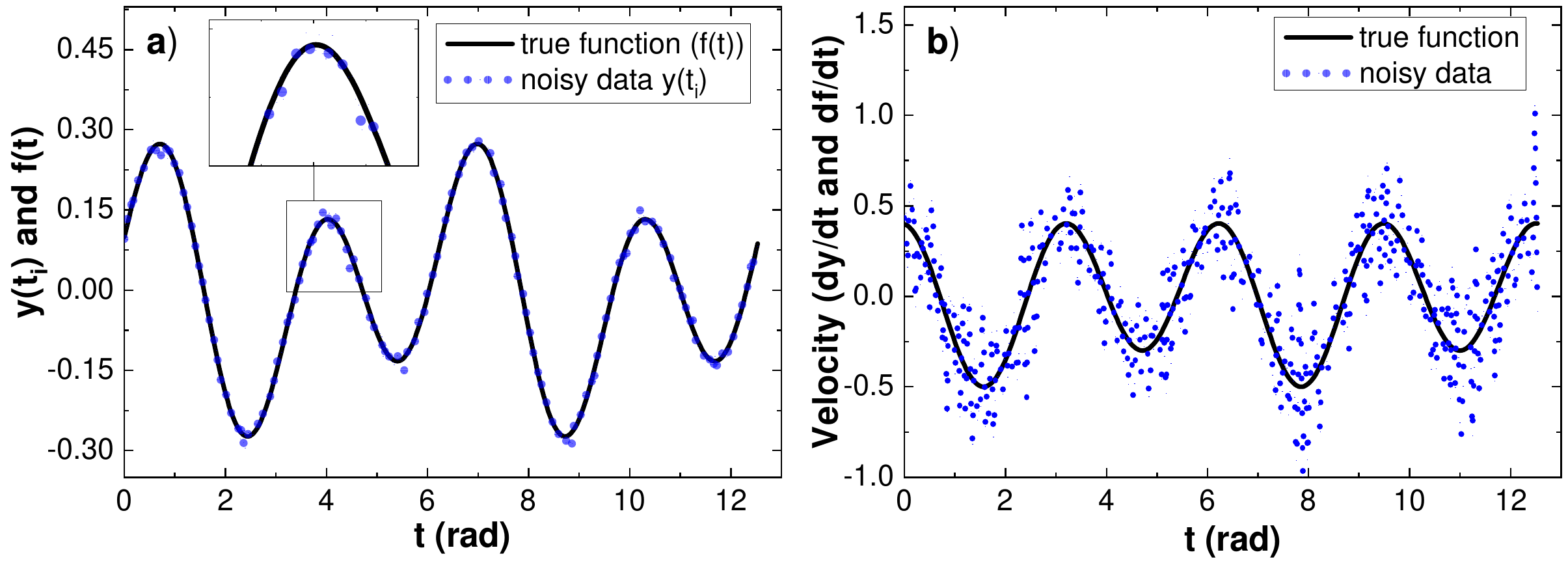}
	\caption{a) True function ($f$) and noisy data ($y_i$) as a function of time ($t$) with a signal-to-noise ratio (SNR) of 25 dB, b) velocity calculated from the noisy data and the true function.}
	\label{fig:TestData}
\end{figure*}

\begin{figure*}[ht]
	\centering
	\includegraphics[width=.8\textwidth]{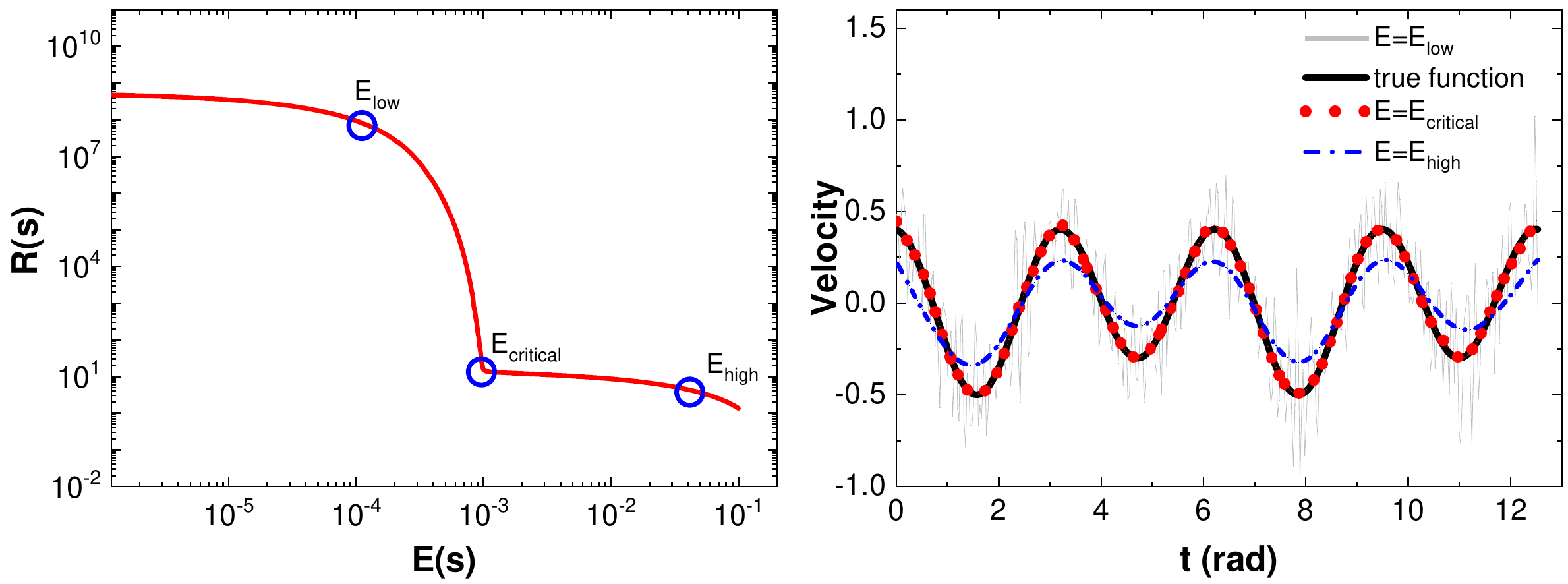}
	\caption{a) Roughness vs Error in log-log scale, roughness and error are calculated based on \ref{eq:Error} and \ref{eq:Rough},  b) velocity calculated from filtered data.}
	\label{fig:TestDatAnalysis}
\end{figure*}

Evaluating particle velocity from the displacement measurements is essential to understanding the physics of materials subjected to shock compression . However, calculating the derivatives from the experimentally measured displacement is nontrivial due to noise. The error in the derivative is typically an order of magnitude higher than the noise in the measurements and is highly susceptible to small errors in the data. Thus, to accurately infer the particle velocity (first time-derivative of displacement), a smoothing spline is generally fit to the displacement vs. time data. The smoothing spline, characterized by its error in relation to the original data and by its roughness, is a piece-wise polynomial fit to the data that can capture any functional form \cite{epps_impulse_2010, de_boor_practical_1978,osullivan_discussion_1985}. The error, $E$, and roughness, $R$, associated with a smoothing spline, $s(t)$, and dataset, ($t_i , y_i$), are given by:

\begin{align}
E(s)&=\sum_{i=1}^{N}\left[y_{i}-s\left(t_{i}\right)\right]^{2} \label{eq:Error} \\
R(s)&=\sum_{i=1}^{N}\left[\frac{d^{3} s\left(t_{i}\right)}{d t^{3}}\right]^{2} \label{eq:Rough}
\end{align}

Determining the optimal smoothing spline fit involves a trade off between the total error of the spline, $E(s)$, and the roughness of the spline, $R(s)$. For example, a spline with zero error would pass through all the (noisy) data points and thus have the highest roughness. A spline with zero roughness would be a straight line of regression through all the data points and thus involve the highest error. 

The majority of literature employs the `de Boor formulation' \cite{de_boor_practical_1978} (terminology adopted from Epps \cite{epps_impulse_2010}) to determine the optimum smoothing spline for the data. The de Boor formulation involves minimizing the weighted sum of the error and roughness measures by an appropriate choice for the weight (smoothing parameter). An equivalent approach can be found in the ‘Reinsch formulation’\cite{reinsch_smoothing_1971} (terminology adopted from Epps \cite{epps_impulse_2010}), which involves finding the smoothing spline with the least roughness but with its error below a chosen threshold. The Reinsch formulation, as implemented in Epps \cite{epps_impulse_2010}, is said to overcome smoothing inadequacies that can result from the de Boor formulation. Epps \cite{epps_impulse_2010} used the Reinsch formulation to construct a plot of minimum roughness vs. critical error tolerance, which was then used to identify the optimal smoothing spline for the data. This procedure is best suited for extracting the velocity history for the region of interest as it is amenable to automation and can operate upon a large number of datasets. This procedure was adopted in the current work and is illustrated next. 

Consider a general sine-cosine signal: $f(t)=\mathrm{A} \sin (t)+\mathrm{B} \cos (t)$ with an added noise $\tilde{\epsilon}(t)$, whose Signal-to-Noise-Ratio (SNR) $=25 \mathrm{~dB}$ and is sampled at $N$ data points. We thus obtain a test dataset $\left(t_{i}, y_{i}\right)$, where $y\left(t_{i}\right)=f\left(t_{i}\right)+\tilde{\epsilon}\left(t_{i}\right)$. The noisy data $y(t_i)$ and the true function $f(t)$ are illustrated in Fig. \ref{fig:TestData}\hyperref[fig:TestData]{a}, and their corresponding derivatives $\left(d f / d t\textnormal{ and } d y_{i} / d t\right)$ are shown in Fig. \ref{fig:TestData}\hyperref[fig:TestData]{b}. It is noted that that the noise in the derivative is significantly larger than the noise in the original data. An outline of the procedure \cite{epps_impulse_2010,reinsch_smoothing_1971} to obtain the smoothing spline to best approximate the function $f(t)$ is described next. The MATLAB \cite{noauthor_matlab_2021} function \textit{spaps} implements the Reisnch formulation (discussed above) to output a minimum roughness smoothing spline for a given noisy dataset and specified maximum error tolerance. The order of the smoothing spline (cubic, quintic, etc.) can also be specified in the function. The noisy dataset $\left(t_{i}, y_{i}\right)$ and a certain maximum error tolerance $E$ are provided to the \textit{spaps} function, which then outputs the smoothed spline datapoints $s\left(t_{i}\right)$. The smoothing spline $s\left(t_{i}\right)$ has the minimum possible roughness $R(s)$ for the given error tolerance of $E(s)$. $E(s)$ and $R(s)$ can be computed using Eqs. (\ref{eq:Error}) and (\ref{eq:Rough}), respectively. By varying the $E(s)$, which is provided to the \textit{spaps} function, various smoothing splines with different $R(s)$ can be obtained. The log-log plot of $R(s)$ vs. $E(s)$, as shown in Fig. \ref{fig:TestDatAnalysis}\hyperref[fig:TestDatAnalysis]{a}, is then analyzed to obtain the critical error tolerance $E_{\textnormal {critical}}$, which is the solution to the de Boor formulation for the optimal smoothing spline. The critical point in the $R(s)$ vs. $E(s)$ plot is the point corresponding to the sudden change in curvature. A plot of the first derivatives of smoothing splines corresponding to error tolerances of $E_{\textnormal{low}}$, $E_{\textnormal{critical}}$ and $E_{\textnormal{high}}$ are shown in Fig. \ref{fig:TestDatAnalysis}\hyperref[fig:TestDatAnalysis]{b} alongside the derivative of the true function, $f(t)$. It can be seen in Fig. \ref{fig:TestDatAnalysis}\hyperref[fig:TestDatAnalysis]{b} that the minimum roughness smoothing spline for the dataset, $y({t_i})$, with a specified error tolerance of $E_{\textnormal{critical }}$ provides an excellent estimate of the first derivative, which closely matches that of the true function. When $E=E_{\text {high}}$, the data is over smoothed, while for $E=E_{\text {low}}$, the data is under smoothed, which will introduce large error while calculating the derivative. This procedure would work only for datasets with low to moderate noise. For datasets with high level of noise, a clear kink in the $R(s)$ vs. $E(s)$ cannot be observed. Further details regarding the procedure and theory used here can be found in Epps \cite{epps_impulse_2010}.

\bibliographystyle{apsrev4-1}
\bibliography{References.bib}

\end{document}